\documentclass[jhep,aps,preprint,groupedaddress,superscriptaddress,nofootinbib,a4paper,preprint]{revtex4}
\usepackage{amsfonts}
\usepackage{amsmath}    
\usepackage{amssymb}
\usepackage[english]{babel}
\usepackage{bbding}
\usepackage{bm}
\usepackage{bbm}
\usepackage{cancel}
\usepackage{color} 
\usepackage{comment}
\usepackage[mathcal]{euscript}
\usepackage{float}
\usepackage{graphics}%
\usepackage{graphicx}   
\usepackage{hyperref} 
\usepackage[latin1]{inputenc}
\usepackage{latexsym}
\usepackage{mathrsfs} 
\usepackage{pifont}
\usepackage{soul}
\usepackage{subfigure}  
\usepackage{tikz}
\usepackage[normalem]{ulem}
\usepackage{verbatim}  

\usetikzlibrary{decorations.pathmorphing}

\begin{document}%

\begin{flushright}
YITP-21-02
\end{flushright}
\title{Inner horizon instability and\\ {the unstable cores of} regular black holes}
\author{Ra\'ul Carballo-Rubio}
\affiliation{
Florida Space Institute, University of Central Florida, 
 12354 Research Parkway, Partnership 1, Orlando, FL, USA}
\author{Francesco Di Filippo}
\affiliation{
Center for Gravitational Physics, Yukawa Institute for Theoretical Physics, Kyoto University, Kyoto 606-8502, Japan}

\affiliation{
SISSA - International School for Advanced Studies, Via Bonomea 265, 34136 Trieste, Italy
}
\author{Stefano Liberati}
\affiliation{
SISSA - International School for Advanced Studies, Via Bonomea 265, 34136 Trieste, Italy
}
\affiliation{
IFPU - Institute for Fundamental Physics of the Universe, Via Beirut 2, 34014 Trieste, Italy
}
\affiliation{
INFN Sezione di Trieste, Via Valerio 2, 34127 Trieste, Italy
}
\author{Costantino Pacilio}
\affiliation{Dipartimento di Fisica, ``Sapienza" Universit\`a di Roma \& Sezione INFN Roma1, Piazzale Aldo Moro 5, 00185, Roma, Italy}
\author{Matt Visser}

\affiliation{
School of Mathematics and Statistics, Victoria University of Wellington, PO Box 600, Wellington 6140, New Zealand
}

\begin{abstract}
Regular black holes with nonsingular cores have been considered in several approaches to quantum gravity, and as agnostic frameworks to address the singularity problem and Hawking's information paradox. While in a recent work we argued that the inner core is destabilized by linear perturbations, opposite claims were raised that regular black holes have in fact stable cores. To reconcile these arguments, we discuss a generalization of the geometrical framework, originally applied to Reissner--Nordtsr\"om black holes by Ori, and show that regular black holes have an exponentially growing Misner--Sharp mass at the inner horizon. This result can be taken as an indication that stable nonsingular black hole spacetimes are not the definitive endpoint of a quantum gravity regularization mechanism, and that nonperturbative backreation effects must be taken into account in order to provide a consistent description of the quantum-gravitational endpoint of gravitational stellar collapse.
\end{abstract}

\maketitle
\section{Introduction}
%
Singularity theorems \cite{Hawking1967,Hawking:1970} demonstrate that,  within the framework of standard classical general relativity, singular black holes are unavoidably formed as the end-state of gravitational collapse \cite{Penrose1964}. Observational tests of black hole spacetimes coming from the detection of gravitational waves emitted by binary black hole mergers \cite{Abbott2016,Abbott2016b,Abbott2017,Abbott2017c,TheLIGOScientific2017} are so far in perfect agreement with the predictions of classical general relativity. Nonetheless, there is still room for alternatives to the classical black holes of general relativity \cite{Carballo-Rubio2018a,Cardoso:2019rvt}, a consideration that becomes more pertinent given that it is reasonable to assume that singularities will be tamed once quantum gravity effects are taken into account \cite{Carballo-Rubio2019b,Carballo-Rubio2019a}.

A conservative approach consists of replacing the singularity with a regular core, without necessarily introducing substantial long-range modifications to the geometry \cite{Dymnikova2001}. Regular black holes emerge in several approaches to quantum gravity such as asymptotic safety \cite{Bonanno2000,Platania:2019} and loop quantum gravity \cite{Modesto:2008im, Rovelli2014}, while they have also been proposed as an agnostic framework to address Hawking's information paradox \cite{hawking1976breakdown,Hayward2006}.

It can be shown that regular black holes, even when they are non-rotating, possess an inner horizon besides the more conventional trapping horizon \cite{Dymnikova2001,Carballo-Rubio2019b,Carballo-Rubio2019a}. It is known that inner horizons in general relativity are generically unstable under small perturbations and give rise to the so called mass-inflation instability \cite{Poisson1989,Ori:1991zz,Hamilton:2008zz}, a fact which is not necessarily problematic given that general relativistic black holes are known to be singular. On the other hand, this opens to the possibility that the inner horizons of regular black holes are also unstable. If confirmed, this would make their physical justification less straightforward, while at the same time it would call for increased theoretical efforts to produce consistent regular models.

In fact, such inner-core instabilities have been derived in the context of a particular model of regular black hole \cite{Brown2011,Bertipagani:2020awe}. Recently, we produced arguments supporting the generic conclusion that regular black hole cores are unstable \cite{Carballo-Rubio:2018pmi}. While the latter analysis captures the main physical ingredients, it considers a somewhat idealized scenario consisting of two non-interacting null shells on top of a static background. In contrast to these results, a recent work \cite{Bonanno:2020fgp} --- which is based on a single perturbing null shell, and takes the backreaction of the metric into account --- extends Ori's work for Reissner--Nordstr\"om black holes \cite{Ori:1991zz} to regular spacetimes and claims that regular black holes have stable cores. This claim contradicts the result found under the same assumptions for the specific regular black hole considered in the Appendix of \cite{Brown2011}.

The aim of the present work is to clarify the issue of regular black hole instabilities, by a reanalysis of the contradictory results in \cite{Carballo-Rubio:2018pmi} and \cite{Bonanno:2020fgp}. In particular, we revisit the extension of Ori's model to a generic spherically symmetric metric. By expressing the equations of motion in terms of the Misner--Sharp mass, which is a physically motivated quantity \cite{misner1964relativistic,Faraoni:2020mdf}, we are able to derive a key equation for its evolution in terms of the initial perturbation. We conduct both an analytical and a numerical study, and we show that a \emph{generic} regular black hole has an unstable core. Therefore, our work makes the conclusions of \cite{Carballo-Rubio:2018pmi} robust, while in contrast it rejects the claims in \cite{Bonanno:2020fgp}.

The paper is organized as follows. In sec.~\ref{sec:ori} we present an extended version of Ori's perturbation model to generic spherically symmetric metrics, and we derive the equation governing the growth of the Misner--Sharp mass. 
In sec.~\ref{sec:instability} we show how this equation generically predicts an indefinite growth for both singular and regular black holes, and we discuss the functional form of the growth rate.
In sec.~\ref{sec:num} we numerically solve the differential equations governing the growth of instability for several specific black hole spacetimes, showing that the solutions are in agreement with the analytical treatment. In sec.~\ref{sec:cosm-const} we analyze how the presence of a non-zero cosmological constant affects the conclusions of the previous sections. 
Concluding remarks are made in sec.~\ref{sec:conclusions}.

\section{{Generalized} Ori model}\label{sec:ori}
In \cite{Ori:1991zz} Ori developed a model to describe the dynamical evolution in the neighborhood of the inner horizon of a Reissner--Nordstr\"om black hole under linear perturbations, in order to explicitly analyze the singularity generated by the well-known mass-inflation instability. Ori's model can be easily {generalized} to any (singular or non{-}singular) spacetime with a metric of the form
\begin{equation}
    \label{eq:metric}
    ds^2 = -f(v,r)dv^2+{2}dvdr+r^2d\Omega^2\,.
\end{equation}
Most of the derivation can be worked out only using the parameterization $f(v,r)=1-2M(v,r)/r$ in terms of the Misner--Sharp mass $M(v,r)$. At this point in the discussion we do not need to assume any specific functional form for the Misner--Sharp mass, but only that its radial dependence is such that the geometry contains an inner horizon: this is equivalent to assuming that $f(v,r)$ has an even number of zeroes, and that all its dependence on $v$ enters through a single-variable function $m(v)$ such that
\begin{equation}
M(v,r)=M(m(v),r),
\end{equation}
and $m(v)$ coincides with the value of the Misner--Sharp mass at large values of $r$, namely
\begin{equation}
    m(v)=\lim_{r\rightarrow+\infty}M(v,r).
\end{equation}
Well-known examples include the Vaidya--Reissner--Nordstr\"om spacetime in which $M(m(v),r)=m(v)-e^2/2r$, or the Hayward spacetime in which $M(m(v),r)=m(v)r^3/(r^3+2m(v)\ell^2)$ \cite{Hayward2006}.

\begin{figure}[h]
\includegraphics[scale=0.8]{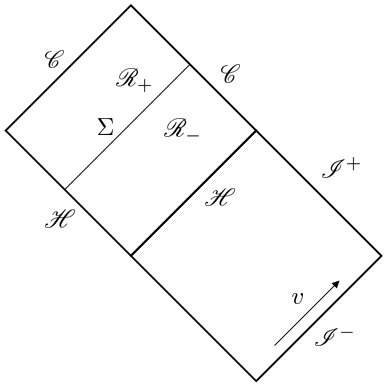}
\caption{Relevant sections of the Penrose diagram of a {regular} black hole. The shell $\Sigma$ divides the spacetime in two regions $\mathscr{R}_-$ and $\mathscr{R}_+$. $\mathscr{H}$ denotes the event horizon while $\mathscr{C}$ denotes the Cauchy horizon. }
\label{fig:diagram}
\end{figure}

Following \cite{Ori:1991zz}, we now consider a dynamical model of a null shell $\Sigma$ which divides the {black hole} into two subregions $\mathcal{R}_-$ and $\mathcal{R}_+$, such that $\mathcal{R}_-$ is on the same side of $\mathscr{I}^{-}$, as shown by the Penrose diagram in fig.~\ref{fig:diagram}.
Therefore we can parametrize $\mathcal{R}_-$ by the same advanced null coordinate $v_-$ of $\mathscr{I}^-$, while we shall choose a distinct null coordinate $v_+$ to parametrize $\mathcal{R}_+$. By continuity, the radial coordinate $r$ is the same on both regions, while the mass function $m(v)$ depends on the region, so that we shall distinguish $m_-(v_-)$ and $m_+(v_+)$.

It can be shown that the matching conditions imply that both $v_-$ and $v_+$ can be expressed \emph{on $\Sigma$} in terms of a single affine coordinate $\lambda$ \cite{Barrabes1991}. We will define $\lambda$ in such a way that it is negative and $\lambda\to0$ when $v_-\to\infty$. Additionally, the radius of the shell can be expressed as a function $R(\lambda)$ of the affine parameter.

The functional form of $m_-(v_-)$ is dictated by Price's law \cite{Price1971,Price1972} to be
\begin{equation}
\label{eq:price}
m_-(v_-)= m_0+\delta m=m_0-\frac{\beta}{v_-^p}\,,\quad p\geq11\,,
\end{equation}
where $m_0=m_-(v\to\infty)$. When $v_-\to\infty$, the shell crosses the corresponding inner horizon at $r_0$, which is implicitly defined by
\begin{subequations}
 \label{eq:r0}
 \begin{align}
  & \lim_{\lambda\to0}R=r_0\,,\\
  & \lim_{\lambda\to0}f_-(\lambda,R)=0\,.
 \end{align}
\end{subequations}
Here $R\equiv R(\lambda)$ is the radius of the shell {and $f_-(\lambda,R)\equiv f(M(m_-(\lambda),R),R)$}. It is convenient to define the surface gravity in $\mathcal{R}_-$ at $\lambda=0$ as
\begin{equation}
 \label{eq:k0}
 \kappa_0=\frac{1}{2}\lim_{\lambda\to0}\left.\frac{\partial f_-(\lambda,r)}{\partial r}\right|_{r=R(\lambda)}\,.
\end{equation}

We now set up the basic ingredients for the discussion of the generalized Ori model. It can be shown that {the matching conditions} on $\Sigma$ in eqs.~(7)--(9) of \cite{Ori:1991zz} generalize, respectively, to
\begin{subequations}
\label{eq:ori}
\begin{align}
    & z{_i} R'=\frac{R}{2}f{_i}(v{_i},R)\label{eq:ori:1}\,,\\
    & v{_i}(\lambda)=\int^\lambda d\lambda\frac{R}{z{_i}}\,,\label{eq:ori:2}\\
    & z{_i}(\lambda)=Z{_i}+\frac{1}{2}\int^\lambda d\lambda\left(f{_i}(\lambda,R)+R\frac{\partial f{_i}}{\partial r}(\lambda,R)\right)\,,\label{eq:ori:3}
\end{align}
\end{subequations}
where $z{_i}=R/v{_i}'${, the index $i$ takes two values $+$ and $-$ (for the two regions $\mathcal{R}_+$ and $\mathcal{R}_-$, respectively),} and a prime denotes differentiation with respect to~$\lambda$. The quantity $Z{_i}$ is an integration constant, while we omitted a similar integration constant in \eqref{eq:ori:2} because it is irrelevant. These equations are valid on both sides of the ingoing null shell. Now, in region $\mathcal{R}_-$, we can expand eq.~\eqref{eq:ori:3} around $\lambda=0$ to obtain 
\begin{equation}
    \label{eq:ori:4}
    z_-(\lambda)= Z_--r_0|\kappa_0|\lambda+\mathcal{O}(\lambda^2)\,.
\end{equation}
However, given that $\lim_{\lambda\to0}v_-=\infty$, it follows from eq.~\eqref{eq:ori:2} that $Z_-=0$, otherwise $v_-$ would tend to a constant proportional to $1/Z_-$. Hereafter, for simplicity, we will use $v$ in place of $v_-$; therefore
\begin{equation}
    \label{eq:ori:5}
    v(\lambda)=-\frac{1}{|\kappa_0|}\ln{|\lambda|}+\mathcal{O}(\lambda^0)\,.
\end{equation}
Notice that the derivation of \eqref{eq:ori:4} and \eqref{eq:ori:5} proceeded exactly as in reference \cite{Ori:1991zz}. Furthermore, eq.~\eqref{eq:ori:1} in region $\mathcal{R}_-$ can be written as a differential equation in $v$ as
\begin{equation}
 \label{eq:rdiff:1}
 \frac{dR(v)}{dv}=\frac{1}{2}f_-({m_-(v)},R(v)).
\end{equation}
{We can perform a Taylor expansion of $f_-$ around the values $m_-=m_0$ and $R=r_0$ in the limit $v\to\infty$ to obtain}
\begin{equation}\label{eq:10r}
 \frac{d\delta R(v)}{dv}=-\frac{A}{r_0}\delta m(v)-|\kappa_0|\delta R(v),
\end{equation}
where we have written ${R(v)=}r_0+\delta R(v)$ and $A=\partial M_-/\partial m_-|_{{m_-=m_0,r=r_0}}$, and we neglected terms of order $\mathcal{O}(\delta m^2, \delta m\delta R, \delta R^2)$. Taking into account the functional form of $\delta m(v)$ in eq.~\eqref{eq:price}, the general solution to the above differential equation is given by
%
\begin{equation}
\label{eq:rsol:1}
\begin{split}
 \delta R(v)&={c_1} e^{-|\kappa_0|v}+\frac{A\beta}{|\kappa_0|r_0v^p}\sum_{k=0}^\infty \left[\frac{(p+k-1)!}{(p-1)!}\frac{1}{|\kappa_0|^k v^k}\right]\\
&=\frac{A\beta}{|\kappa_0|r_0v^p}\left\{1+\frac{p}{|\kappa_0|v}+\frac{p(p+1)}{|\kappa_0|^2v^2}+\mathcal{O}(v^{-3})\right\}\,.
 \end{split}
\end{equation}

Let us now consider {the evolution of the metric functions in region $\mathcal{R}_+$ on its boundary $\Sigma$, hence as functions of $\lambda$ only (alternatively, $v$)}. One can manipulate eqs. \eqref{eq:ori:1} and \eqref{eq:ori:3} {for $i=+$} to obtain
\begin{equation}
 \frac{df_+{(\lambda,R)}}{d\lambda}=R'\left.\frac{\partial f_+{(\lambda,r)}}{\partial r}\right|_{r=R}+\frac{R''}{R'}f_+{(\lambda,R)},
\end{equation}
or equivalently, as a differential equation in $v$,
\begin{widetext}
\begin{equation}
\begin{split}
\label{eq:mdiff:1}
\frac{dM_+{(v,R)}}{dv}&=R_{,v}\left.\frac{\partial M_+{(v,r)}}{\partial r}\right|_{r=R}-\left(\frac{\lambda_{,vv}}{\lambda_{,v}}-\frac{R_{,vv}}{R_{,v}}\right)\left(M_+{(v,R)}-\frac{R}{2}\right)\\
&=R_{,v}\left.\frac{\partial M_+{(v,r)}}{\partial r}\right|_{r=R}+\left(|\kappa_0|-\frac{p+1}{v}{+\mathcal{O}(v^{-2})}\right)\left(M_+{(v,R)}-\frac{R}{2}\right)\,.
\end{split}
\end{equation}
\end{widetext}
In the equation above the $v$ subindices denote differentiation with respect to this variable. This is the main equation that we derive in the paper. In the next section we are going to comment its solutions and show that it predicts unstable cores for regular black holes.
\section{Instability of regular black holes}
\label{sec:instability}
In the case of Vaidya--Reissner--Nordstr\"om spacetime
\begin{equation}
  \left.\frac{\partial M_+{(v,r)}}{\partial r}\right|_{r=R}=\frac{e^2}{2{R}^2},
\end{equation}
so the first term of the right-hand side {of eq.~\eqref{eq:mdiff:1}} does not depend on $M_+$ and the equation {displays} (exponentially) inflating solutions with leading behaviour
\begin{equation}
\label{eq:msol:1}
 M_+{(v,R(v))}\propto\frac{e^{|\kappa_0|v}}{v^{p+1}}\,.
\end{equation}
{This is the solution obtained by Ori \cite{Ori:1991zz}.} More generally, this result is recovered unchanged for a wider family of geometries satisfying the linear ansatz $M_+(v,r)=g_1(r)m_+(v)+g_2(r)${, which includes for instance the Bardeen metric for a regular black hole \cite{Bardeen1968}}.

However, if $\partial M_+/\partial r$ is not linear in $M_+$, {the differential equation \eqref{eq:mdiff:1} becomes nonlinear and the last term on the right-hand side} may drive the evolution at late times, {thus modifying} the exponential behavior associated with mass-inflation. Crucially, not all the metrics of interest are covered by a linear ansatz. For instance, in the case of a {regular black hole described by the Hayward metric \cite{Hayward2006}, the functional form of $M(m,r)=mr^3/(r^3+2m\ell^2)$ implies that}
\begin{equation}\label{eq:hayder}
\frac{\partial M}{\partial r}=\frac{6\ell^2}{r^4}M^2.
\end{equation}
We will be more general and consider the case where $\partial M_+/\partial r$ is polynomial in $M_+$, i.e., $\partial M_+/\partial r\propto M_+^n$. We then obtain
\begin{widetext}
\begin{equation}
\label{eq:final}
 \frac{dM_+{\left(v,R(v)\right)}}{dv}{=}\left(|\kappa_0|-\frac{p+1}{v}{+\mathcal{O}\left(v^{-2}\right)}\right)\left(M_+{\left(v,R(v)\right)}-\frac{r_0}{2}\right){-\frac{\gamma}{v^{p+1}}M_+^n\left(v,R(v)\right)}
\end{equation}
\end{widetext}
for some given constants $\gamma$ 
and $n$, which depend on the specific metric under consideration. The behaviour of the solution depends on the sign of $\gamma$. It is possible to show that, for $\gamma<0$, $M_+{\left(v,R(v)\right)}$ diverges for a finite value of $v$. On the other hand, for $\gamma>0$ we can solve this differential equation in different regimes. At early times, the first term on the right-hand side dominates the 
dynamics, and leads to an exponential growth of $M_+{\left(v,R(v)\right)}$, up to a critical time $v_0$ for which 
\begin{equation}\label{eq:asymp_M2a}
 M_+^{n-1}(v_0,R(v_0)){\approx}\frac{|\kappa_0|}{\gamma}v_0^{p+1}.
\end{equation} 
From that point on, the first and the second term of the right-hand side become comparable and the growth is polynomial rather than exponential,
\begin{equation}\label{eq:asymp_M2}
 M_+^{n-1}\left(v,R(v)\right)\propto v^{p+1},\qquad {v\gg v_0}.
\end{equation} 
As a concrete example, we consider Hayward's metric and the relation \eqref{eq:hayder}, from which it follows that
\begin{equation}
\label{eq:ex_hay}
\frac{dM_+{\left(v,R(v)\right)}}{dv}\simeq|\kappa_0|M_+{\left(v,R(v)\right)}-\gamma\frac{M_+^2{\left(v,R(v)\right)}}{v^{p+1}}-\frac{r_0|\kappa_0|}{2},
\end{equation}
{where the symbol $\simeq$ means that we are neglecting subdominant terms in the $v\rightarrow\infty$ limit, and}
\begin{equation}
\label{eq:gamma-Hay}
\gamma=\frac{6p\beta\ell^2}{|\kappa_0|}\frac{r_0}{\left(2m_0\ell^2+r_0^3\right)^2}.
\end{equation}
By neglecting the last term on the right hand side, the solution of \eqref{eq:ex_hay} is given by
\begin{equation}
M_+(v,R(v)){\approx}\frac{e^{|\kappa_0|v} v^p}{c_1 v^p-(-v)^p|\kappa|^{p-1}\gamma\Gamma(-p,-|\kappa_0|v)}.
\end{equation}
Expanding the incomplete Gamma function at late times we obtain
\begin{equation}
\label{eq:transition:point}
M_+(v,R(v)){\approx}\frac{|\kappa_0|}{\gamma}v^{p+1},
\end{equation}
in perfect agreement with eq.~\eqref{eq:asymp_M2}. Therefore, at late times, the growth rate of the instability is slowed down, and becomes polynomial rather than exponential.

In the next section, we are going to integrate the perturbative equations of motions numerically for several regular black hole metrics usually considered in the literature. We will show that, for all the metrics under consideration, $M_+$ grows indefinitely with $v$. While the functional form (exponential or  polynomial) of the growth rate depends on the specific form of the metric, the fact that inflation occurs appears to be a generic phenomenon. Therefore, we find evidence that regular black holes have unstable cores; while, in principle, we cannot exclude the existence of a fine-tuned metric avoiding this conclusion, we found no concrete example of such behaviour and we expect it to be non-generic.

Before closing this section, we provide an argument for the fact that the transition from the exponential to the polynomial growth of $M_+$, when present, might occur beyond the regime of validity of the linear analysis. For concreteness, we refer to Hayward's metric. If the corrections to the Schwarzschild metric come from quantum regularization effects, it is reasonable to expect that the regularization parameter satisfies $\ell\ll m_0$. Therefore, from eq.s~\eqref{eq:gamma-Hay} and \eqref{eq:transition:point}, we find that the transition occurs when
\begin{equation}
\label{eq:M2-pol}
\begin{split}
M_+(v,R(v))&\approx\frac{|\kappa_0|^2\left( 2m_0\ell^2+r_0^3 \right)^2	}{r_0\ell^2}\frac{v_0^{p+1}}{6p\beta }\\
&\sim\frac{m_0^2}{\ell}\frac{v_0^{p+1}}{6p\beta},
\end{split}
\end{equation} 
where we used $r_0\sim|\kappa{_0}|^{-1}\sim\ell$. In turn, this implies
\begin{equation}
\frac{M_+(v,R(v))}{m_0}\sim\frac{m_0}{\ell}\frac{v_0^{p+1}}{6p\beta}\gg1\,,
\end{equation}
namely, at the transition point the value of $M_+$ is already exponentially large w.r.t.~the initial value $m_0$. Let us stress that this argument reinforces the case for mass-inflation as it argues that, in the situations of physical interest, the dominant behaviour will be effectively exponential.
\section{Numerical analysis}\label{sec:num}
In this section, we provide explicit numerical solutions for the perturbative equations of motion for several choices of the black hole metric which have been considered in the literature. We can integrate eqs.~\eqref{eq:rdiff:1} and \eqref{eq:mdiff:1} for the variables $R(v)$ and $M_+(v,R(v))$. However, we will follow an alternative route and integrate the equations for the variables $R(v)$ and $m_+\left(v\right)$. While the main advantage of the first method is the use of the physically motivated Misner--Sharp mass as one of the variables, the second method will clarify some subtleties that arise in the interpretation of the mass-inflation phenomenon when the metric is expressed in terms of the unphysical variable $m_+(v)$.

An equation for $m_+(v)$ can be derived following the treatment in \cite{Bonanno:2020fgp}, apart for the correction of a mistake, which explains the opposite conclusions we reach in our work. The starting point is the junction condition at the shell $\Sigma$ \cite{Barrabes1991}
\begin{equation}
\label{eq:junc}
\left[T_{\mu}\,^\nu s^\mu s_\nu\right]=0,
\end{equation}
where $T_{\mu}\,^\nu$ is the effective stress energy tensor obtained by imposing Einstein's equations on both regions $\mathcal{R}_{\{+,-\}}$ of the spacetime, and 
\begin{equation}
s^\mu=\left(2/f_\pm,1,0,0\right)
\end{equation}
is the outgoing null normal to the shell. The relevant components of the stress energy tensor are
\begin{subequations}
\begin{align}
&T_v\,^v=-4\pi r^2\frac{\partial M}{\partial r},\\
&T_v\,^r=4\pi r^2\frac{\partial M}{\partial v},\\
&T_r\,^r=-4\pi r^2\frac{\partial M}{\partial v}.
\end{align}
\end{subequations}
Then eq.~\eqref{eq:junc} becomes
\begin{equation}
\label{eq:jbonanno1}
\left.\frac{1}{f_+^2}\frac{\partial M_+(v_+,r)}{\partial v_+}\right|_{r=R(v_+)}=\left.\frac{1}{f_-^2}\frac{\partial M_-(v_-,r)}{\partial v_-}\right|_{r=R(v_-)}.
\end{equation}
We can eliminate the $v_+$ dependence from this equation noting that, along a null trajectory,
\begin{equation}
\frac{dv_+}{dv_-}=\frac{f_-}{f_+}
\end{equation}
and expressing everything in terms of $v\equiv v_-$
\begin{equation}
\label{eq:bonanno2}
\left. \frac{1}{f_+}\frac{\partial M_+}{\partial v}\right|_{r=R(v)}=\left.\frac{1}{f_-}\frac{\partial M_-}{\partial v}\right|_{r=R(v)}\,.
\end{equation}
Given that the evolution of $m_-(v)$ is dictated by \eqref{eq:price} and that $M_+$ is implicitly a function of $m_+$, the system of eq.s \eqref{eq:bonanno2} and \eqref{eq:rdiff:1} can be solved in terms of $m_+(v)$ and $R(v)$.
In ref.~\cite{Bonanno:2020fgp} eq.~\eqref{eq:bonanno2} was interpreted as if the partial derivatives in $v$ are evaluated \emph{after} imposing $r=R(v)$ on $\Sigma$. This is inconsistent with how \eqref{eq:bonanno2} is derived and, crucially, it is the reason why \cite{Bonanno:2020fgp} reaches the conclusion that regular black holes have stable cores.

\begin{figure*}
\includegraphics[width=.45\linewidth]{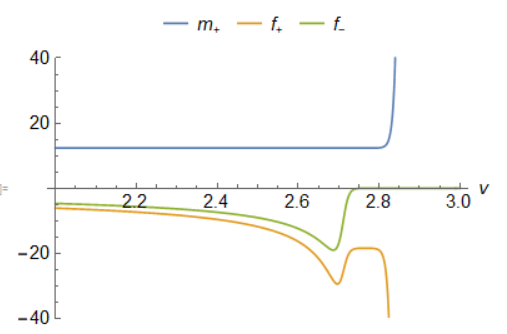}
\includegraphics[width=.45\linewidth]{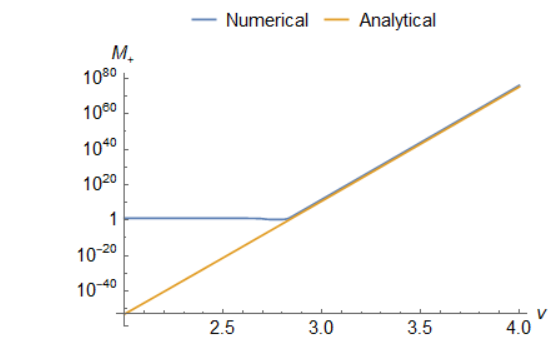}
\caption{Left: Numerical evolution of the mass parameter $m_+$ and of the metric functions $f_\pm$ of a Reissner--Nordstr\"om black hole. Right: Comparison between the numerical and the analytical evolution of the Misner--Sharp mass $M_+$. In both plots we have used the parameters $\beta=1$, $p=12$, $m_0=10$, $e=5$ with the initial conditions $R(v=1)=5$ and $m_+(v=1)=m_0+1$.}\label{Fig:RN}
\end{figure*}

\begin{figure*}
\includegraphics[width=.45\linewidth]{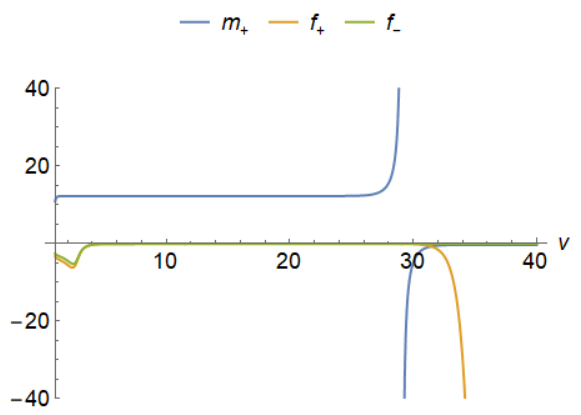}
\includegraphics[width=.45\linewidth]{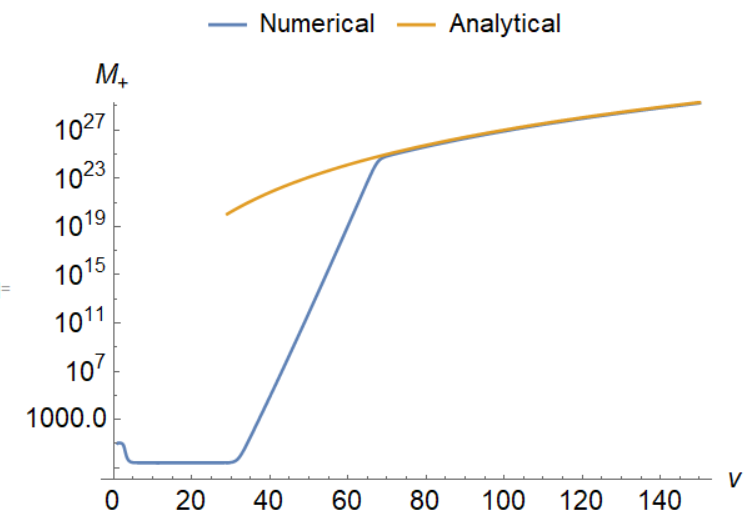}
\caption{Left: Numerical evolution of the mass parameter $m_+$ and of the metric functions $f_\pm$ of a Hayward regular black hole. Right: Numerical evolution of the Misner--Sharp mass. In both plots we picked the parameters $\beta=1$, $p=12$, $m_0=10$, $\ell=0.5$ with the initial conditions $R(v=1)=5$ and $m_+(v=1)=m_0+1$.}
\label{Fig:Hayward}
\end{figure*}

\begin{figure*}
\includegraphics[width=.45\linewidth]{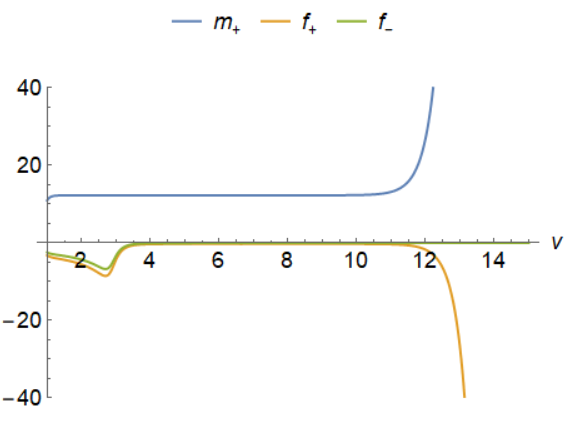}
\includegraphics[width=.45\linewidth]{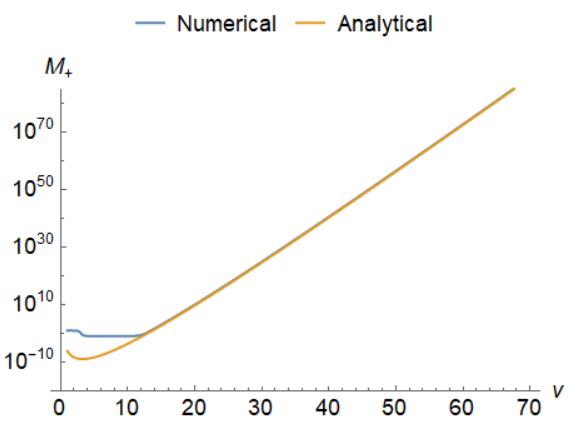}
\caption{Left: Numerical evolution of the mass parameter $m_+$ and of the metric functions $f_\pm$ of a Bardeen regular black hole. Right: Numerical evolution of the Misner--Sharp mass. 
In both plots we picked the parameters $\beta=1$, $p=12$, $m_0=10$, $\ell=1$ with the initial conditions $R(v=1)=5$ and $m_+(v=1)=m_0+1$.}\label{Fig:Bardeen}
\end{figure*}

\begin{figure*}
\includegraphics[width=.45\linewidth]{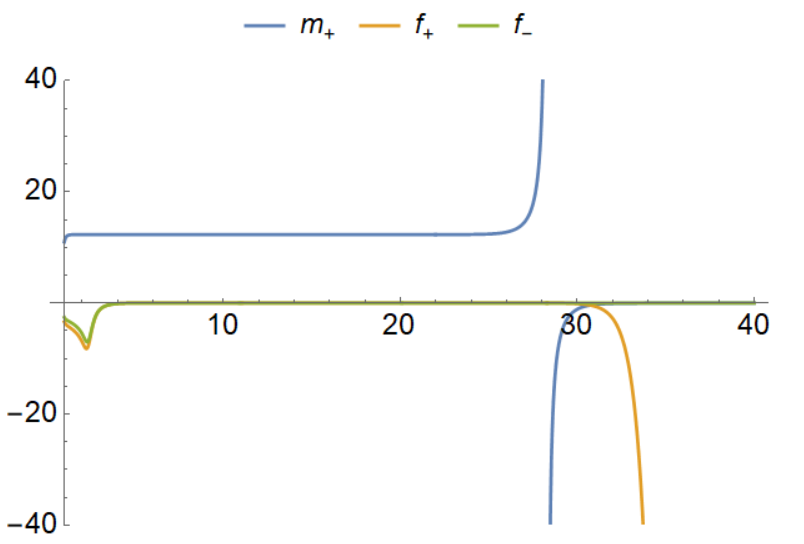}
\includegraphics[width=.45\linewidth]{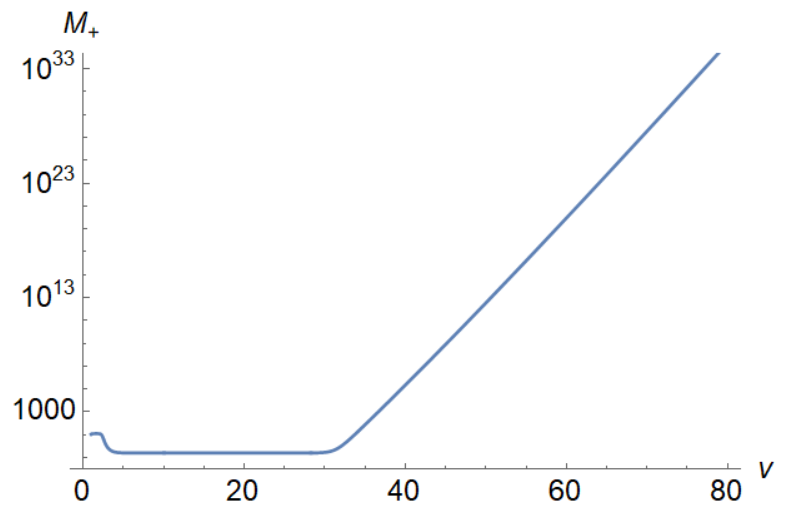}
\caption{Left: Numerical evolution of the mass parameter $m_+$ and of the metric functions $f_\pm$ of a Dymnikova regular black hole. Right: Numerical evolution of the Misner--Sharp mass. In both plots we picked the parameters $\beta=1$, $p=12$, $m_0=10$, $\ell=0.5$ with the initial conditions $R(v=1)=5$ and $m_+(v=1)=m_0+1$.}\label{Fig:Dyminikova}
\end{figure*}

We shall now present some explicit examples of solutions for the metric coefficients displayed in table \ref{tab:metrics}. We will also show the corresponding evolution of $M_+$, as deduced from the solutions of $m_+$.
\begin{table}[]
    \centering
    \begin{tabular}{|c|c|}
         metric &  $M(v,r)$\\
         \hline\hline
         Reissner-Nordstr\"om & $m(v)-\frac{e^2}{2r}$\\
         [1mm]
         Hayward \cite{Hayward2006}& $\frac{m(v)r^3}{r^3+2m(v)\ell^2}$\\
         [1mm]
         Bardeen \cite{Bardeen1968}& $\frac{m(v)r^3}{\left(r^2+l^2\right)^{3/2}}$\\
         [1mm]
         Dymnikova \cite{Dymnikova1992}& $m(v)\left(1-e^{-r^3/\ell^2m(v)}\right)$
    \end{tabular}
    \caption{Black hole parametrizations used for the numerical integrations in sec.~\ref{sec:num}.}
    \label{tab:metrics}
\end{table}
\paragraph{Reissner-Nordstr\"om metric}
First, as a sanity check, we integrate the system for the Reissner-Nordstr\"om black hole, for which eq.~\eqref{eq:bonanno2} becomes 
\begin{equation}
\label{eq:RN}
\frac{m_+'(v)}{R^2 -2 R\, m_+(v)+e^2}=\frac{m_-'(v)}{R^2 -2 R\, m_-(v)+e^2}.
\end{equation}
The results of the numerical integration are plotted in fig.~\ref{Fig:RN}. As expected, the system develops a mass inflation instability and the Misner--Sharp mass diverges exponentially, see fig.~\ref{Fig:RN} (right). At the same time, as it is also apparent from the functional form of the metric, the functions $m_+$ and $f_+$ exhibit corresponding divergences, see fig.~\ref{Fig:RN} (left).
\paragraph{Hayward metric}
The case of Hayward metric is less trivial. Indeed, as shown in fig.~\ref{Fig:Hayward} (left), the integration for $m_+$ breaks down at a finite value of $v$, for which $m_+$ diverges. However, the integration can be continued from the opposite side of the discontinuity, where $m_+$ is negative and approaches $m_+\to-r_0^3/(2\ell^2)$ asymptotically for $v\to\infty$. We can see from the form of the metric that the limiting value of $m_+$ is such that $M_+$ diverges, as expected.  While it might seem odd that the variable $m_+$ has a discontinuity and becomes negative, the reader should take in mind that $m_+$ is not a physical variable and so this behaviour is an artifact of the parametrization. On the other hand, the physical variable $M_+$ is well behaved, as shown in fig.~\ref{Fig:Hayward} (right). Indeed, $M_+$ follows a first phase of exponential expansion, followed by a phase of polynomial growth. This is in complete agreement with the analysis of the previous section.
\paragraph{Bardeen metric}
The functional form of the Bardeen metric is such that $M_+$ is linear in $m_+$. Therefore, from the previous section, we expect an exponential inflation of $M_+$. This is validated by the numerical results in fig.~\ref{Fig:Bardeen} (right). A direct comparison with fig.~\ref{Fig:RN} also shows that $m_+$ and $f_+$ behave similarly to the Reissner-Nordstr\"om case.
\paragraph{Dynmikova metric}
As our final example, we consider the regular black hole metric proposed by Dymnikova \cite{Dymnikova1992}. Since in this case $\partial M_+/\partial r$ is not a polynomial, we cannot apply the reasoning of the previous section and we have no prior expectations. A direct numerical integration (fig.~\ref{Fig:Dyminikova}) shows that $M_+$ diverges exponentially. On the other hand, $m_+$ presents a change-of-sign discontinuity analogous to the case of the Hayward metric (fig.~\ref{Fig:Hayward}). Again, we stress that this is an artifact of the parametrization. 

\section{Role of the cosmological constant}
\label{sec:cosm-const}
So far we ignored the role of the cosmological constant. Intuitively, this is justified by the fact that core instabilities happen at the inner horizon, where the energy density is Planckian and the role of the cosmological constant is completely negligible. However, in the presence of a non-zero  cosmological constant, Price's law \eqref{eq:price} is modified --- it exhibits an exponential decay of the perturbations at late time \cite{barreto1997,dyatlov2012}
\begin{equation}
\label{eq:cosm-decay}
m_-{(v)=m_0+\delta m(v)}=m_0-\alpha e^{-\omega_I v},
\end{equation}
where $\omega_I>0$ is the imaginary part of the least damped quasinormal mode of oscillation of the black hole. 

If the cosmological constant is small, the fall-off of the perturbations follow the polynomial Price's law of eq.~\eqref{eq:price} for a very long time and, therefore, the eventual exponential fall-off in eq.~\eqref{eq:cosm-decay} can be ignored. This is certainly the case in our universe, in which therefore core instabilities will manifest well before the cosmological constant could play any significant role.

For theoretical completeness, we investigate the scenario in which a non-zero cosmological constant does play a significant role, to inspect if it can be fine-tuned to suppress the core instability. We parallel the discussion in sec.s~\ref{sec:ori} and \ref{sec:instability}. First, using \eqref{eq:cosm-decay}, the solution to eq.~\eqref{eq:rdiff:1} becomes
\begin{equation}
\delta R(v)=c_1e^{-|\kappa_0|v}+\frac{A\alpha}{r_0(|\kappa_0|-\omega_I)}e^{-\omega_Iv}
\end{equation}
where $c_1$ is an integration constant. This, in turn, results in the following equation for the evolution of $M_+$
\begin{widetext}
\begin{equation}
\frac{dM_+{\left(v,R(v)\right)}}{dv}\simeq-f_+A\alpha\omega_Ie^{-\omega_Iv}\left(2|\kappa_0|c_1e^{-|\kappa_0|v}+\frac{2\omega_IA\alpha e^{-\omega_Iv}}{r_0(|\kappa_0|-\omega_I)}\right)^{-1}+\frac{f_-}{2}{\left.\frac{\partial M_+(v,r)}{\partial r}\right|_{r=R(v)}}.
\end{equation}
\end{widetext}
It is convenient to analyze separately the cases $|\kappa_0|>\omega_I$ and $|\kappa_0|<\omega_I$. When $|\kappa_0|>\omega_I$ we have
\begin{equation}
\label{eq:m+:lambda}
\frac{dM_+{\left(v,R(v)\right)}}{dv}{\simeq}\frac{r_0(|\kappa_0|-\omega_I)\left(2M_+{\left(v,R(v)\right)}/r_0-1\right)}{2}+\frac{f_-}{2}{\left.\frac{\partial M_+(v,r)}{\partial r}\right|_{r=R(v)}}.
\end{equation}
Following the same reasoning as in sec.~\ref{sec:instability}, if $\partial_r M_+$ is linear in $M_+$ then the last term becomes subdominant and $M_+$ grows exponentially with leading behaviour
\begin{equation}
M_+{\left(v,R(v)\right)}\propto e^{(|\kappa_0|-\omega_I)v}.
\end{equation}
This is analogous to eq.~\eqref{eq:msol:1}, in which the polynomial fall-off in $v$ is replaced by the exponential decay of \eqref{eq:cosm-decay}. On the other hand, at variance with the flat case, if $\partial_r M_+$ is polynomial in $M_+$, i.e., $\partial_r M_+\propto M_+^n$, we still find that the inflation is governed by an exponential growth 
\begin{equation}
    M_+{\left(v,R(v)\right)}\propto e^{\omega_Iv/(n-1)}\,.
\end{equation}

When $|\kappa_0|<\omega_I$, eq.~\eqref{eq:m+:lambda} becomes
\begin{equation}
\frac{d M_+{\left(v,R(v)\right)}}{dv}{\simeq}\frac{\left(2M_+{\left(v,R(v)\right)}/r{_0}-1\right)}{2|\kappa_0|}A\alpha e^{-(\omega_I-|\kappa|_0)v}-2|\kappa_0|e^{-|\kappa_0|v}{\left.\frac{\partial M_+(v,r)}{\partial r}\right|_{r=R(v)}}
\end{equation}
and we see that, irrespective of the specific form of the metric, the asymptotic behaviour is regular. Thus, under this conditions, it might be theoretically possible to avoid mass-inflation under linear perturbation (although, as we already explained, this is not going to occur in our universe). Notice, however, that $|\kappa_0|<\omega_I$ is a property of the single black hole under consideration and, in general, it will not be satisfied for all black holes in a gravitational theory \cite{Cardoso2018b,Dias2018,Hod2018,Cardoso:2018nvb,Bhattacharjee:2016zof,Bhattacharjee:2020}; therefore, the avoidance of mass-inflation will not be a generic feature.
\section{Conclusions}
\label{sec:conclusions}
In this paper, we have reanalyzed the issue of the inner horizon instability for regular black hole spacetimes, in light of recent contradictory results \cite{Carballo-Rubio:2018pmi,Bonanno:2020fgp}. We extended Ori's perturbative model to a generic spherically symmetric spacetime, showing that mass inflation at the inner horizon is a robust and very general prediction.
In particular, while the occurrence of mass-inflation might be obscured by the choice of parameterization, the phenomenon is evident when the equations are expressed in terms of the Misner--Sharp mass.

We found that the Misner--Sharp mass always experiences a period of exponential instability which can either go on indefinitely (e.g., with the Bardeen metric) or be followed by a period of polynomial growth (e.g., with the Hayward metric). In both cases, the growth does not stop. Additionally, we argued that the onset of the polynomial instability is likely to occur when the approximation of linear perturbations has already broken down. Therefore, exponential mass-inflation appears to be the leading instability mechanism of regular black hole cores.

When a nonvanishing cosmological constant is allowed, we found that in principle the inner core might become stable. However, this occurs case-by-case and it is not a generic feature of the theory.
Furthermore, even when the conditions for stability are realized, they require the late-time tail of the perturbation (which is sensitive to the presence of the cosmological constant) to be physically relevant, which is not going to be the case in our universe due to the very small value of the cosmological constant.

We stress that the instability of the inner horizon does not necessarily imply the formation of a physical singularity. What we have shown is that an initial linear perturbation has a huge effect on the geometry and leads to an unbounded growth of the Misner--Sharp mass. While in general relativity this usually implies the formation of a singularity, we might expect that, in a full theory of quantum gravity, backreaction drives the geometry to a different class of non-singular spacetimes \cite{Carballo-Rubio2019a,Barcelo:2020mjw}. Therefore, in order to address the endpoint of the inner core instability, a geometrical analysis is not informative enough.

A rigorous approach would require to specify the dynamical field equations of the theory leading to the formation of regular black holes. Possible quantum gravity frameworks explored in the literature are asymptotic safety and loop quantum gravity. Indeed, within these theories, several studies have argued that the singularity of general relativistic black holes is replaced by a nonsingular core with an inner horizon~\cite{Bonanno2000,Platania:2019,Held:2019xde,Modesto:2008im,Alesci:2011wn,Rovelli2014}. While certainly a demanding task, we hope that our work serves as a stimulus for the quantum gravity community to address the effects of nonperturbative backreaction on regular spacetimes in a theoretically consistent framework.

\begin{acknowledgments}
\noindent
FDF acknowledges financial support by Japan Society for the Promotion of
Science Grants-in-Aid for Scientific Research No.~17H06359.\\
SL acknowledges funding from the Italian Ministry of Education and  Scientific Research (MIUR)  under the grant  PRIN MIUR 2017-MB8AEZ.\\
CP acknowledges: financial support provided under the European Union's H2020 ERC, Starting Grant agreement no.~DarkGRA--757480; support under the MIUR PRIN and FARE programmes (GW-NEXT, CUP:~B84I20000100001); support from the Amaldi Research Center funded by the MIUR program ``Dipartimento di Eccellenza'' (CUP: B81I18001170001). \\
MV was supported by the Marsden Fund, via a grant administered by the Royal Society of New Zealand.
\end{acknowledgments}
\bibliographystyle{utphysics}
\bibliography{refs}
\end{document}